\documentclass[12pt]{article}
\makeatletter
\newcommand{\LyX}{L\kern-.1667em\lower.25em
\hbox{Y}\kern-.125emX\spacefactor1000}
\def\mb#1{\setbox0=\hbox{$#1$}\kern-.025em\copy0\kern-\wd0
\kern-0.05em\copy0\kern-\wd0\kern-.025em\raise.0233em\box0}

\makeatother

\begin{document}

\title{Particle Aggregation in a \\
Turbulent Keplerian Flow}

\author{A. Bracco, P.H. Chavanis, A. Provenzale \\
\sl{Istituto di Cosmogeofisica, Corso Fiume  10133 Torino} \\
E.A. Spiegel \\
\sl{Astronomy Department, Columbia University}}

\maketitle

\abstract{In the problem of planetary formation one seeks a mechanism to
gather small dust particles together into larger solid
objects.  Here we describe a scenario in which turbulence mediates this
process by aggregating particles into anticyclonic regions. If, as our
simulations suggest, anticyclonic vortices form as long-lived coherent
structures, the process becomes more powerful because such vortices
trap particles effectively.   Even if the turbulence is decaying,
following the upheaval that formed the disk, there is enough time to make
the dust distribution quite lumpy.}

\section{Introduction}
\label{sec_intro}

To rationalize the distribution and the motions of the planets and their
satellites in the solar system, Kant$^{26}$ and Laplace$^{29}$
separately introduced forms of the nebular hypothesis, a
conjecture that the solar system formed from a flattened gas cloud or
disk.   Their prescience has been confirmed: disks do occur in many
places in the universe.  Since much of the interesting behavior in disks
has been ascribed to turbulence, a problem motivated by them is
appropriate in a volume dedicated to R.H. Kraichnan.  That problem is to
understand how the planets formed in the primitive solar nebula.

Stars condense from an interstellar medium consisting mostly of gas with
an admixture of solid particles called interstellar dust.  Both
observational and numerical studies suggest that, as the central star
contracts, it leaves around it material that contains a good share of the
initial angular momentum of the whole system.  In this nebula, the
centrifugal force balances the stellar gravity in the radial direction
and a disk is formed.  As the dust settles toward the midplane of this
disk it is somehow accumulated into protoplanetary objects.

The general picture has been reviewed by Lissauer$^{30}$
who has provided an extensive bibliography. Other summaries exist
(such as Ref.49) and there are a number of introductory
expositions (for example Ref.11).  Two main possibilities have
been considered.  If the disk is not turbulent,
solid dust particles settle into a very thin layer and then can
accumulate into larger solid objects by gravitational instability$^{39,24}$.
On the other hand, if the disk sustains
turbulence, it has been thought that one must rely on a slower process of
adhesion or growth by two-body collisions.  We examine here an alternative
possibility in which turbulence in the early history of a
protoplanetary disk leads to a rapid aggregation of particles by
anomalous diffusion. Even if the
turbulence  is decaying, there is time for the particles to aggregate
significantly.

The idea that the formation of planets in the Kant-Laplace disks, was
somehow mediated by the interaction of turbulence and rotation was
adumbrated by von Weizs\"aker.$^{48}$  Subsequently, experiments
showed how a field of vortices of opposite senses can produce
inhomogeneous particle distributions.$^{12,13}$  The
experimental evidence that anticyclonic vortices (whose vorticity is in
the sense counter to that of the prevailing global motions) can trap
particles is now extensive.$^{45,47}$

An application of
the mechanism of anomalous diffusion by anticyclonic vortices is to the
formation of planets,$^{6,44}$ 
as seen in several numerical studies
of Langrangian orbits in prescribed vorticity fields.$^{36}$
In this paper, we consider
flows of the kind that would arise in a Keplerian disk and follow the
aggregation process they produce.  Our purpose is to show how this
aggregation process can work despite some objections to the notion
(to be discussed
below) that protoplanetary disks can sustain turbulence.$^{4,5}$
Without turbulence, it is unlikely that vortices on disks
form at all.

Our way around this impasse is to note that the initial
aggregation of dust could take
place in the decaying turbulence that was likely to have occured in
the formation of the nebula.  Once this has happened, the usual sequel
of gravitational clumping can be presumed to take over.  We shall present
simulations to show how anticyclonic vortices form from sufficiently
large initial perturbations of Keplerian disks
(see also Refs.10,2,41), just as in earlier simulations
with mild shears,$^{32,21}$ and strongly promote
the aggregation of dust.  The question we then investigate is whether the
vortices have time before they die away to strongly aggregate the
particles. 

In the next section, we write down the equations of motion to set the
notation in order.  We turn then to a brief summary of relevant
aspects of disks, after which we exhibit results of simulations of a
Keplerian flow.  These show how long-lived anticyclonic vortices form
from initial finite disturbances.  Then, from a study of the Lagrangian
orbits of dust in this flow, we show how an initially homogeneous
distribution of particles is redistributed into strong local
accumulations of particles.

\section{Equations of Motion}
\label{sec_hydro}

To concentrate on vortex dynamics, we restrict ourselves to
incompressible flow with \begin{equation}
\nabla \cdot {\bf v}  = 0 \ .  \label{inc} \end{equation}
We also assume constant viscosity so that the equation of motion is
\begin{equation}
{{ D {\bf v}}\over {Dt}}=-{1\over \rho}\nabla p
-\nabla \Phi +\nu \Delta {\bf v} \ ,
\label{NS}
\end{equation}
where $\Phi=-GM/r$ is the gravitational potential of the central
mass, the protosun.

The key quantity here is the vorticity, for which the
evolution equation is
\begin{equation}
{D{\mb{\omega}}\over Dt} ={\mb{\omega}} \cdot \nabla {\bf v} +
\nu\Delta {\mb{\omega}} \ .
\label{vorticity}
\end{equation}

In a disk, the large scale flow is nearly two-dimensional and we adopt
this simplification here. Then we may express ${\bf v}$ in terms
of a stream function $\psi$ such that, in Cartesian coordinates, the
velocity field is given by \begin{equation}
\left( -{\partial \psi \over \partial y} \ , {\partial \psi \over
\partial x}
\right) \ . \label{psi}
\end{equation}
The vorticity then has only one component, $\omega$, and it is related to
the streamfunction by
\begin{equation}
\omega=\nabla^2\psi \; .
\label{zeta}
\end{equation}
Hence equation (\ref{vorticity}) may be written as
\begin{equation}
{{\partial \omega} \over {\partial t}} + J(\psi,\omega) = \nu \Delta\omega
\label{2d}
\end{equation}
where $J(\psi,\omega)=\partial_x \psi \partial_y \omega -  \partial_x
\omega \partial_y \psi$.

In the absence of dissipation ($\nu=0$), equation (\ref{2d}) admits an
infinite number of conserved quantities, two of which are quadratic
invariants; these are the kinetic energy $E = 1/2 \int{(\nabla
\psi)^2 dxdy}$ and the enstrophy $Z = 1/2 \int{(\nabla^2 \psi)^2 dxdy}$.
The conservation of these two quantities in the inviscid limit is what
makes two-dimensional turbulence so different from three-dimensional
turbulence at large Reynolds numbers.$^{7,27}$
In two-dimensional turbulence there is a direct cascade of enstrophy from
large to small scales and an inverse cascade from small to large scales
of kinetic energy.$^{17,38,28}$

For freely decaying, barotropic turbulence in shear-free environments,
intense long-lived vorticity concentrations are observed to form after an
energy dissipation time.  These coherent vortices are characterized by a
broad distribution of size and circulation and they contain most of the
energy and the enstrophy of the system.$^{33,34}$
The first question we take up here is whether the strong Keplerian shears
of accretion disks prevent this well-documented behaviour.

\section{Astrophysical Disks}
\label{sec-blah}

The purpose of this section is to provide a general idea about the
background for the fluid processes that we study in the next
sections.  This is called for in a controversial subject such as planet
formation since many readers of this journal may not have
previously encountered some of the issues raised. On the other hand, the
occurence of disks of many sizes throughout the universe is so well
documented that our discussion can safely start from the assumption that
one of these once surrounded the protosun.  We shall not go into the
processes by which disks are thought to form but consider only the
simplest case of a disk that is axisymmetric in the large and symmetric
about the midplane $z = 0$.  In the radial direction, the principal
support against the gravitational pull of the protosun is provided by
centrifugal force while, in the $z$-direction, the support is by a
pressure gradient.  If the ratio of these two support forces is large, the
disk is quite thin, as we shall presume here.

Let the rotational speed at distance $r$ from the central protosun be $v$.
The centrifugal balance condition is \begin{equation}
v^2=2GM/r \ , \label{bal} \end{equation}
where $M$ is the mass
of the protosun.  This describes a Keplerian flow that is stable to small
perturbations since it has no inflection points and the Rayleigh
discriminant, \begin{equation}
\Delta=r^{-3}d(r^2\Omega)^2/dr \ , \label{dis} \end{equation}
is positive.  In astronomy, where this stability condition is often met,
it is custumary to introduce the epicyclic frequency
$\kappa=\sqrt{\Delta}$, whose role in the dynamics is not wholly different
from the buoyancy frequency of stably stratified fluids.  Since the disk
is thin, we may identify $r$ with the radius in cylindrical coordinates.
In our two-dimensional simulations, we retain this meaning of $r$ and the
form $-GM/r^2$ for the gravitational acceleration.

If the flow remains laminar, there will be a slow viscous transport of
angular momentum outward that allows a weak inward spiralling of the
material onto the protosun.  The two-dimensional version of such motion
has been studied in the context of drag reduction theory$^{8}$
where it is thought that a slow inflow can inhibit boundary layer
separation and reduce the drag on airfoils.  This process is too slow for
astrophysical purposes when it is laminar, but its turbulent analogue has
been invoked for the case of astrophysical disks,$^{31}$ which are
called accretion disks because of this inflow.  In the turbulent case, it
is suggested, the dissipation rate could be high enough to render some
disks quite bright.  But can we assume that disks are turbulent?

Until recently, many took it for granted that Keplerian disks were
turbulent because the Reynolds number in a disk is literally astronomical.
On account of (\ref{bal}) and the relatively small viscosity, it was
often assumed$^{15}$ that ``The successive rings of gas
in the medium will therefore have motions relative to one another, and
turbulence will ensue.'' But simulations by Balbus and collaborators,$^3$
among others, have not shown nonlinear instabilities in
Keplerian disks at the respectable Reynolds numbers they achieved.

Apparently, the circular Keplerian flow is different from plane Couette
flow, where nonlinear instability is known to occur, and indeed does so in
the calculations run by Balbus {\it et al}.  Nevertheless, their
conclusion is not universally accepted$^{23}$ and there is an
issue of principle to be decided here.  One technique that may be
decisive is the method of energy stability which may prove helpful in
deciding whether there can be sustained turbulence in nonmagnetic
Keplerian disks.

On the other hand, for the typical astrophysical conditions, the matter in
disks is ionized and is likely to be magnetized.  A prevalent magnetic
field can destabilize a shear flow that would 
otherwise have been stable$^{16,43,18}$
and Balbus and Hawley$^{5}$
have introduced this notion into disk theory to great
effect to justify the assumption that turbulence does occur in accretion
disks.  Once started, this magnetoturbulence may be expected to produce a
dynamo and to be self-sustaining along with the turbulence.   In this way,
the study of disk structure can be said to have been reduced to a
previouly unsolved problem.  The literature on this problem is of course
quite large, in keeping with its importance in astrophysics.

By contrast, in the case of the disk that is supposed to have spawned the
solar system, it is thought that the matter was too cool to be ionized
and so the flux was not frozen in Ref.46.  We cannot then call on
magnetic effects to render the solar nebula turbulent.  The problem of
whether such a disk is turbulent is therefore open, and there are at
this time no real grounds for invoking turbulence.

However that issue is decided, it is likely that most would accept the
assumption that the initial conditions in the disk were turbulent, and
this is sufficient for our purposes, even if that initial turbulence were
decaying.  Indeed, our results indicate that the decay lasts many rotation
periods, and that is ample time for the vortical aggregation process
we describe here to make good headway.

\section{Simulations with Keplerian Shear }
\label{sec_simu}

If disks are turbulent, they might be expected to follow the pattern of
other rotating turbulent fluids$^{9,25,33,42,32,21,22}$
and form coherent vortices that last many turnaround times.  Even for
turbulent disks, this possibility is not generally accepted because the
shears are so strong in Keplerian flows.  In this section we study this
problem to see whether vortices may form and survive for long times in
Keplerian shear given a finite initial disturbance.  A positive result
would imply that even if the protoplanetary disk were nonlinearly stable,
it would have produced vortices early in its lifetime since it is likely
to have been formed in a turbulent state.  In this section we find that
for initial perturbations above some threshold, long-lived vortices do
form, though they eventually decay.  Then, in the next section, we
investigate whether the vortices have enough time in their decay periods
to have a significant effect on the particle distribution in a disk and so
contribute the formation of planets.

To see whether vortices can form in a Keplerian shear, we perform
numerical simulations of an incompressible, two-dimensional flow
driven with a shear like that derived from (7).  Into this flow,  we
introduce sizable perturbations such as one might expect in an {\it
initially} turbulent situation and follow the formation of vortices that
indeed survive for many rotation periods of the disk.   This approach is
like that used previously$^{34}$ to simulate vortex formation in
decaying turbulence and it permits us to study the effect of strong shear
in this problem, which turns out to be that anticyclonic vortices are
heavily favored.

\subsection{Technical details}
\label{sec_setup}
The basic quantity that we compute is the vorticity,
which we split into two parts, a specified (regularized) Keplerian flow
and a disturbance to the basic flow.  Thus we write,
\begin{equation}
\omega=\omega_{K}(r)+\zeta({\bf r},t) \ ,
\label{decomposition}
\end{equation}
where $\zeta$ is the dynamical quantity of interest.

The imposed background Keplerian velocity profile is, in regularized form,
taken to be
\begin{equation}
v_{K}(r)={K\over \sqrt{r}}\left\{ 1-\exp [-({r\over r_{0}})^{3/2}]\right\}
\label{velocity}
\end{equation}
where $K$ and $r_{0}$ are adjustable parameters.  For $r\gg r_{0}$, a
conventional Keplerian profile $K/\sqrt{r}$ is recovered while for
$r\ll r_{0}$, the velocity field reduces to a solid rotation $K
r/r_{0}^{3/2}$.  The behavior at small $r$ is introduced to avoid the
singularity at the origin associated with a point mass.
This profile has a vorticity field,
$\omega_{K}={1\over r}{\partial\over\partial r}(r v_{K})$,
given by
\begin{equation}
\omega_{K}(r)= {K\over 2 r^{3/2}} \left\{
1-\exp [-({r\over r_{0}})^{3/2}]+ 3\left({r \over
r_{0}}\right)^{3/2}\exp[-({r\over r_{0}})^{3/2}] \right\}
\label{vorticitykep}
\end{equation}
For $r\ll r_{0}$, $\omega_{K}\simeq
2K/ r_{0}^{3/2}$ and for $r\gg r_{0}$, $\omega_{K}\simeq
K / 2 r^{3/2}$.

Our simulations are based on a standard pseudo-spectral code with
2/3 dealiasing and resolution $512\times 512$ grid points in a square box
with periodic boundary conditions $\lbrack -\pi,\pi\rbrack\times\lbrack
-\pi,\pi\rbrack$.  In the simulations discussed below, we have used the
value $r_{0}=0.123$, which corresponds to $10$ grid points in the region
of rigid rotation.  We also adopt $K=2.07$, which gives the mean kinetic
energy of the Keplerian shear the value $E_{K}=0.5$.  With these choices,
we ensure a Keplerian profile with $\omega_{K}\propto 1 / r^{3/2}$ for the
underlying vorticity and velocity fields in the range $0.3<r< 2$ (see
figure 1).  For $r<0.3$, the disk rotates rigidly and for $r>2$ the pure
Keplerian profile is altered by the periodicity of the domain and the
interaction with disk images from the periodic boundary conditions.  At
$r=\pi/6$, the angular velocity is $2\omega_{K}\simeq 5.46$, which gives a
typical rotation time $T_{K}\simeq 1.15$.  The simulations have been run
up to a total time $T=20$.

When we introduce the decomposition (\ref{decomposition}) into
(\ref{2d}), we obtain an evolution equation for the disturbance vorticity,
$\zeta$:
\begin{equation}
{\partial\zeta\over\partial t}+
J(\varphi,\zeta)=\nu\Delta\zeta+F \label{evolution}
\end{equation}
\begin{equation}
F\equiv -J(\psi_{K},\zeta) - J(\varphi, \omega_K)
\label{F} \end{equation}
where $\varphi$ and $\psi_{K}$ are the streamfunctions
associated with $\zeta$ and $\omega_{K}$ respectively. The quantity
$F$ represents the effect on the disturbance flow of the imposition of
a background Keplerian shear, which satisfies $J(\psi_{K},\omega_{K})=0$.
We have assumed that the dissipation acts only on the disturbance field
$\zeta$ and we have taken $\nu=5\ \times 10^{-5}$.

As initial condition, $\zeta({\bf r},0)$, we select a narrow-band random
vorticity field with energy spectrum$^{34}$ 
\begin{equation}
E(k)=E_0{ { k^n} \over {[(m/n)k_0+k]^{m+n}}} \; .\label{spectrum}
\end{equation}
We have taken $k_{0}=10$, $m=30$ and $n=5$.  The value of $E_0$ is chosen
to control the relative energies of the Keplerian and of the disturbance
flow, as specified below. The Fourier phases are randomly distributed
initially between $0$ and $2\pi$. In order to avoid unrealistic
periodicity effects, the disturbance has been limited to a region of
radius $r_{max}=2$, so that no perturbations occur near the edges of the
simulation box.

\subsection{Formation of anticyclonic vortices}
\label{sec_anticyclones}

In our simulations, when the initial energy of a disturbance is
less than about $10^{-3}$ of the energy in the Keplerian flow, the
vorticity fluctuations are sheared away and the disk returns to its
unperturbed velocity profile. This behavior is consistent with the linear
stability of Keplerian disks. When we begin a calculation with larger
perturbation energies than this, an initially random vorticity field forms
itself into well defined vortices, much as in flows with little or no
shear, except that our vortices are all anticyclonic.  Cyclonic vortices
do not form and, even if they are expressly introduced initially, the
shear quickly destroys them.

Figure 2 shows the vorticity field for a simulation in which the initial
disturbance energy was chosen to be $4 \times 10^{-3} E_K$.  The
vorticity is shown at times $t=0$ (figure 2a), $t=8$ (figure 2b) and
$t=15$ (figure 2c). The presence of anticyclonic vortices is clearly
visible.  Because they last many of their own turnover times, such
vortices are an example of coherent structures.

The initial sizes of the developing anticyclonic vortices are
small, but they merge and grow until they apparently reach a limiting
size. The limiting size of the vortices increases with the distance to the
center of the disk roughly like $r^{1.5}$ out to $r\sim 1.5$, though small
vortices penetrate into the central regions of the disk.
In this system, formation and persistence of coherent vortices is
possible only when the nonlinear term $J(\varphi,\zeta)$ in eq. 
(\ref{evolution}) is larger than dissipation $\nu\Delta\zeta$,
large-scale shear $J(\psi_{K},\zeta)$ and differential rotation 
$J(\varphi, \omega_K)$. For Keplerian disks, the range of scales where
vortices can persist is thus limited from below by dissipation and
from above by shear and differential rotation. Whenever the two limiting
scales become too close, vortices cannot survive. 

For thinking of the solar nebula, we identify a length scale such that
the border of the disk, at $r_{max}=2$, corresponds to $7 AU$. (An
astronomical unit is the mean distance of the earth from the sun.) With
this convention, the biggest vortices of our simulation are located at
typically $5 AU$, at the distance of Jupiter from the center, while the
smallest vortices lie at roughly the earth's distance from the sun.

Once the length scale has been chosen, the time scale is 
determined by Kepler's law of rotation. First we recall that the
rotation period is $T_K \simeq 1.15$ at $r = \pi/6$ in adimensional
units. In physical units, this distance corresponds to about $1.83 AU$.
At this distance, the (physical) revolution time is
$T_K^* \simeq 2.48$ yr.
Physical time is therefore related to 
time in adimensional units by
\begin{equation}
t^* (year)=2.16\ t \,\,\,\,\, .
\label{timescale}
\end{equation}

To give the value of the viscosity coefficient a physical meaning,
we recall that turbulent viscosity in disks is usually expressed
in terms of the dimensionless $\alpha$-parameter$^{40}$ by
\begin{equation}
\nu=\alpha H^{2} \omega_{K}^{*}
\label{alpha}
\end{equation}
where $H$ is the thickness of the disk and 
$\omega_{K}^{*}$
is the Keplerian rotation in the region of interest
(in physical units). 
In the standard model of the
solar nebula, $\alpha$ is in the range $10^{-3}-10^{-2}$ in
the region of Jupiter, assuming 
$H\sim 0.3 AU$. 
This determines
a physical eddy viscosity $\nu\sim 5\times 10^{-5}-5\times 
10^{-4}(AU)^{2}/year$. 
In adimensional
units, it corresponds to $8\times 10^{-6}-8\times 10^{-5}$. 
The viscosity used in the
previous simulations, $\nu=5\times 10^{-5}$, 
lies in this range, indicating that
formation of coherent vortices is indeed possible when the parameters
take the values commonly estimated for the solar
nebula.

\section{The Aggregation Process}
\label{sec_transport}

Anticyclonic vortices trap small particles.$^{47,6,44,36}$
In this section, we
study the effectiveness of this process when an explicit solution of the
equations of motion is used to provide the vortices.  In particular, we
ascertain that the slowly decaying vortices in Keplerian turbulence have
time to produce significant particle aggregation and so initiate planetary
formation before they die away.

\subsection{The Particle Equation}
\label{sec_equations}

We next follow the Langrangian motion of dust particles in the flow
computed in the previous section.  This motion is described in a rigidly
rotating frame with an angular velocity that is approximately the mean
of the Keplerian rotation over the disk, namely $\Omega = 0.617$.

The study of particle motion in a fluid flow is surprisingly 
complicated,$^{35,36}$
especially for fluffy, extended flakes.
But here, we consider the simplest case of dust particles whose sizes
are less than the mean free path in the ambient gas and whose typical
density, $\rho_d$ is much greater than $\rho_g$, the density of the gas
around it.   For this case, the equation of motion of a dust particle
with position ${\bf r}$ is
\begin{equation}
{d^{2}{\bf r}\over dt^{2}}=-2{\bf \Omega}\wedge {d{\bf
r}\over dt} - \gamma\biggl ({d{\bf r}\over dt}-{\bf v}({\bf r},t)\biggr
)+\left(\Omega^{2}-{GM\over r^3}\right)\ {\bf r} \label{particle}
\end{equation}
where ${\bf v}({\bf r}, t)$ is the fluid velocity in the rotating frame
and $\gamma$ is a friction coefficient.  In addition to the gravitational
force on the particle, this equation contains the Coriolis force, the
centrifugal force and a phenomenological drag force produced by the
ambient medium.

For the rarefied conditions of the protoplanetary nebula,
$\gamma\propto \rho_{g}/ a$ where $a$ is the particle radius; this
corresponds to the so-called Epstein regime.$^{30}$  Our
simulations of the particle motion show that the value of $\gamma$ is
important in the aggregation process.  For large $\gamma$, the particles
are rather like tracers and for small $\gamma$ the concentration process
is more rapid.

\subsection{Trapped particles}
\label{sec_trapping}

We start our simulation of the Lagrangian trajectories with a uniform
distribution of particles moving initially at the local fluid velocity.
This leads to the results shown in figure 3 for $\gamma=10$.  The
successive panels display the particle distributions at
times $t=2, 8, 15$ and $20$.
To bring out the effect of vorticity, $\zeta$, in figure 4 we provide
a plot of the time evolution of the average
density of particles in three different regions, namely where $\zeta<
-\zeta_{0}$ (full line), $-\zeta_{0}< \zeta < \zeta_{0}$ (dotted line)
and $\zeta >\zeta_{0}$ (dashed line) with
$\zeta_{0}$ as an estimate of the r.m.s. initial vorticity.
The plotted densities are divided by the
mean density, so that the initial distribution is everywhere unity.
We observe that
the particles are expelled from the cyclonic regions and concentrated
inside the anticyclonic vortices.  Simulations with other values of the
drag parameter $\gamma$ have confirmed this general picture; the particle
concentration process is even more rapid for larger particles associated
with smaller values of $\gamma$. A full study of the role of the
various parameters appearing in eq.(\ref{particle}) is reported in
Ref.44 for a kinematic flow field.

These results are qualitatively consistent with 
earlier empirical work$^{47,12}$
and with theoretical studies in which the
particle paths were followed in an {\it a priori} prescribed flow
containing vortices$^{6,44,14}$.
The quantitative result added here is that decaying vortices in a real
flow derived from the fluid equations have adequate time to aggregate
particles.

Thus, by marching the particle motion along with the simulations that
produce the vortices we have been able to see how turbulent fluctuations
initially present in the disk draw particles into regions of negative
vorticity.  Already in the early evolution of the solar nebula we observe
the development of pockets of dust which fill the disk inhomogeneously
even as the vortices form.  This separation of the dust into different
regions is caused by the cyclone/anticyclone asymmetry, and it is so rapid
that at $t=2$ (a matter of years) the density of the dust particles in the
region $\zeta< -\zeta_{0}$ has already been increased by a factor $2$.
When the vortices do form, they are already surrounded by a large number
of dust particles and so the rate of capture is enhanced.

\section{Discussion}

Rotating cosmic bodies whose atmospheres are turbulent and can be seen
with good resolution display coherent structures --- vortices in the
atmospheres of planets and magnetic flux tubes in the solar atmosphere.
However, many accretion disk theorists, feel that this observation is
not a good indicator of what happens on disks because they feel that the
strong Keplerian shears in accretion disks will destroy vortices.  There
is practically no direct observational evidence on this point, but
indirect evidence does suggest that coherent structures form in the disks
around quasars.$^1$  Our fluid dynamical simulations, (see also
Ref.10) show that Keplerian shears feed anticylonic vortices and
destroy only the cyclonic ones.  For vortices to form however strong
perturbations such as might be provided by turbulence are needed.

It is believed that the progenitor of the solar system was so cool that
is was not ionized and so magnetic flux would not have been frozen in.
Then the powerful magnetic instabilities that render hot disks turbulent
would not operate and there is no well established mechanism for
sustaining turbulence.  Nevertheless, we have argued here that one may
call upon a residual turbulence from the formation throes of the solar
nebula.  The vortices that would have been formed in those initial stages
do seem to have time to act significantly on the particle distribution
before decaying away.   It is even possible that the process might have
involved some of the primordial fields in the collapsing interstellar
matter.  Such details aside, it does seem reasonable to conclude that,
once there is some turbulence, anticylonic vortices form.   This in
itself is an interesting fluid dynamical process that is likely to take
place on conventional accretion disks.  The version of the problem that
we have looked at has features in common with two-dimensional turbulence,
which R.H.  Kraichnan has done much to explicate.$^{27,28}$

Of course, the situation is more complicated in the real astrophysical
contexts than the version we have presented here and a more realistic
version of the theory might be based on shallow layer theory.$^{10}$
Even if if the fluid is compressible, as it is
in astrophysical disks, the thin layer approximation is very close
to the standard shallow water equations,$^{37}$ which also
give rise to vortex formation if strong disturbances are 
introduced.$^{19,20}$
Whether vortices form in disks or not,
large vortical disturbances can by themselves redistribute particles.
This may in itself be enough to initiate planet formation, but a complete
theory would require a more detailed knowledge of how the nebula formed.

In more general terms, the (perhaps) surprising ability of turbulence to
render an initially homogeneous distribution of ambient particles very
inhomogeneous is likely to have consequences in several fields and
deserves further study in the wake of the recent theoretical efforts in
this direction.$^{47,14,36}$  This
problem might also profit from a statistical treatment analogous to
what has been done in the problem of the advection of a passive scalar,
another subject to which R. H. Kraichnan has made a significant
contribution.

\section{Acknowledgments:}  We are grateful to Steve Balbus and Peter
Goldreich who discussed their work with us during the 1998 GFD Summer
Program at the Woods Hole Oceanographic Institution. A letter from
Jean-Paul Zahn clarified his views for us.

\pagebreak
\centerline{{\bf Figure Captions}}

\noindent Figure 1. The background velocity profile used in the
   simulations. The conventional Keplerian profile is plotted as a dashed
   line and the regularized profile is in full line.

\noindent Figure 2a. Initial vorticity field of the perturbed
   two-dimensional Keplerian disk ($t=0$).

\noindent Figure 2b. Vorticity field at $t=8$.

\noindent Figure 2c. Vorticity field at $t=15$.

\noindent Figure 3a. Distribution of the particles in the disk
at time $t=2$
for $\gamma=10$.

\noindent Figure 3b. Distribution of the particles  at $t=8$.

\noindent Figure 3c. Distribution of the particles at $t=15$.

\noindent Figure 3d. Distribution of the particles at $t=20$.

\noindent Figure 4. Evolution of the average density of particles in
different regions of the disk characterized by specific values of the
vorticity: $\zeta< -0.5$ (full line), $-0.5< \zeta < 0.5$
(dotted line) and $\zeta > 0.5$ (dashed line).
\end{document}